\documentclass[12pt]{iopart}
\usepackage{amsfonts} 
\usepackage{graphicx} 
\usepackage[utf8]{inputenc}
\usepackage{xcolor}
\usepackage{hyperref}

\begin{document}

\title[Understanding student challenges with circulation...]{\textcolor{black}{
Understanding student challenges with circulation and Ampère-Maxwell law}}

\author{Álvaro Suárez$^1$, Arturo C Marti$^2$\footnote{Corresponding author}, Kristina Zuza$^3$, Jenaro Guisasola$^4$}

\address{$^1$ Departamento de Física, Consejo de Formación en Educación, Montevideo, Uruguay}
\address{$^2$ Facultad de Ciencias, Universidad de la República, Montevideo, Uruguay}
\address{$^3$ Department of Applied Physics, University of Basque Country, Spain}
\address{$^4$ School of Dual Engineering, Institute of Machine Tools (IMH), Elgoibar, Spain}
\ead{alsua@outlook.com}
\ead{marti@fisica.edu.uy}
\ead{kristina.zuza@ehu.eus}
\ead{jenaro.guisasola@ehu.eus}

\date{\today}

\begin{abstract}
  \textcolor{black}{ We investigated how students apply the concept of
    circulation when faced with problems related to the Ampere-Maxwell
    law.} For this purpose, we designed a metacognitive
  pencil-and-paper question, presented it to 65 students, and analysed
  their responses using phenomenography. We complemented our research
  by conducting interviews with 12 students. The results show that
  students tend to think that they can only use the magnetic field
  circulation to calculate magnetic fields if the curve used is
  symmetric, and that they do not take into account the shape of the
  magnetic field lines when applying it. We also find that some
  students believe that a variable electric field must cross the
  entire Amperian curve in order to apply Ampère-Maxwell’s law and
  find the magnetic field.
\end{abstract}

\maketitle


\section{\label{Intro}Introduction}

Maxwell's equations are the cornerstone of classical electromagnetic
theory. In Physics Education Research (PER), there have been many
studies of the conceptual difficulties that students face in applying
them, with particular attention paid to Gauss's law of the electric
field
\cite{guisasola2003analisis,guisasola2008gauss,singh2006,pepper2012,Li_2017,Li_2018,Campos2023,Hernandez2023}
and Faraday's law
\cite{Bagno1997,Saarelainen_2007,thong2008some,zuza2012,guisasola2013university},
but less research on Ampère's law and, in particular,
Ampère-Maxwell's law.

With regard to Ampère's law, a significant percentage of students use
reasoning based on memorisation and mechanical application of the
information received in class and incorrectly consider that the
magnetic field circulation can always be expressed as the product of
the magnetic field and the length of the curve. They also think that
the magnetic field used in the calculation of the line integral
originates only from the currents crossing a surface bounded by the
curve \cite{guisasola2003analisis,guisasola2008gauss}. Additionally,
they assume that when the net current is null, the magnetic field also
vanishes \cite{guisasola2003analisis,guisasola2008gauss,wallace2010}.
Students often do not use the available information about the magnetic
field to calculate the circulation, nor do they realise that the
magnetic field circulation is a sum of elements $\vec{B} \cdot
\overrightarrow{d l}$ along a closed curve \cite{wallace2010}. They
also confuse the concepts of circulation with magnetic field and
circulation with flux \cite{Campos2023,Hernandez2023}.

In relation to Ampère-Maxwell's law, a recent work addresses whether
students understand the limitations of Ampère's law and how they
understand and apply Ampère-Maxwell's law \cite{suarez2024learning}.
The results of this study show that students struggle to recognise the
limitations of Ampère's law and to identify the current that crosses
a surface bounded by a curve.  It is also revealed that students have
difficulty in recognising when a displacement current appears, as well
as properly relating the magnetic field circulation to the rate at
which the electric field varies.

This work is a continuation of research into students' conceptual
difficulties with Ampère-Maxwell's law \cite{suarez2024learning} and
is part of a wider project with the ultimate aim of developing a
teaching-learning sequence \cite{suarez2024learning,suarez2023}. In
this context, and considering the conceptual difficulties identified
in the previous study \cite{suarez2024learning}, \textcolor{black}{we
  have extended our original research to further explore the various
  qualitative approaches students use when applying magnetic field
  circulation in situations requiring explicit calculation from the
  magnetic field.} Our research question is:
\begin{itemize}
    \item \textcolor{black}{
    In what distinct qualitative ways do students apply magnetic field circulation in the context of Ampère-Maxwell’s law?}
\end{itemize}

To address this research question, we designed a metacognitive
pencil-and-paper question that led students to think about their own
learning.  This question proposes a problem and a possible solution,
so that the students must evaluate whether the solution is correct or
not. To deepen the explanations received, we conducted semi-structured
interviews with several groups of students.

\textcolor{black}{Metacognition refers to an individual's ability to be
  aware of and regulate their own thought processes. It encourages
  individuals to reflect on their thinking, critically analyzing it
  for deeper understanding and continuous improvement
  \cite{Martinez2006}. While metacognition is often defined as
  “thinking about thinking” or “learning about learning,” it
  also involves the use of problem-solving strategies such as
  planning, monitoring, and evaluating \cite{Martinez2006}. This
  approach has been widely recommended for investigating learning
  through understanding and active engagement, as it is broadly
  recognized that developing metacognitive skills is a crucial
  component of the learning process \cite{Mota,Schwartz}. In the
  following sections, we outline the methodological approach and the
  context of the research and design. We then present the results
  obtained, followed by a discussion, and conclude with the
  implications for teaching Ampère-Maxwell's law.}

\section{\label{Métodología} \textcolor{black}{Methodological approach}}

\textcolor{black}{Students' conceptions have been investigated through
  different tasks and in different contexts, and the coherence of
  these conceptions is a key aspect to consider
  \cite{clough1986study}.  This issue raises the need for studies that
  describe the variability in students' conceptions. Phenomenography
  has been proposed and employed as a methodological approach to
  analyse and explain this variability
  \cite{marton1981phenomenography,marton1997booth}.  Here, we adopt a
  phenomenographic approach to explore the various qualitative ways in
  which individuals experience, conceptualize, perceive, and
  understand different phenomena and aspects of the world around them
  \cite{marton1981phenomenography}. Marton and Booth explain that }
\begin{quote}
\textit{\textcolor{black}{“in phenomenography individuals are seen as the bearers of different ways of experiencing a phenomenon and as bearers of fragments of differing ways of experiencing that phenomenon.” \cite[p~114]{marton1997booth}.}}
\end{quote}

\textcolor{black}{In this approach, the description elicited from learners is collective, leaving aside individual voices. Phenomenography examines how diverse ways of perceiving and understanding reality (concepts and associated modes of reasoning) can be organised into categories that describe reality. These categories represent shared patterns of understanding, visible across many individuals, and thus reflect a kind of collective knowledge:}
\begin{quote}
\textit{\textcolor{black}{“The same description categories appear in different situations. The set of categories is thus stable and can be applied, even if individuals “move” from one category to another on different occasions.” \cite[p~195]{marton1981phenomenography}.}}
\end{quote}

\textcolor{black}{
Marton and Booth \cite{marton1997booth} propose certain criteria for creating these categories: (a) each category should be clearly related to the phenomena studied, providing a specific perspective on how they are experienced; (b) the categories should have a hierarchical structure, progressing from simpler to more complex relationships; and (c) the category system should be parsimonious, explained with the fewest categories reasonably possible. Meeting these criteria makes the category system useful both theoretically and pedagogically \cite{trigwell2000phenomenography}. These categories encompass two dimensions: the referential aspect and the structural aspect. The referential aspect pertains to the general meaning that individuals attribute to the phenomenon, reflecting the overarching interpretation that emerges in their experience. In contrast, the structural aspect focuses on how they organize and structure their understanding of the phenomenon. Both aspects are intrinsically linked and, together, facilitate a comprehensive representation of the phenomenon under investigation \cite{MartonPong2005}. }

\textcolor{black}{
In this study, concepts are presented in descriptive categories according to Marton and Booth's criteria. These categories are derived from the data obtained in the questionnaires and interviews, without attempting to force the data into pre-established categories. Each category highlights features that distinguish one concept from another and is organised hierarchically to reflect increasing levels of understanding. This hierarchy of descriptive categories shows the relationships between conceptions and provides a basis for teaching and assessment decisions \cite{Guisasola2023PR}.}

\section{\label{} \textcolor{black}{Research and design framework}}

\textcolor{black}{Our objective is to identify the different problem-solving approaches used by university students when applying the concept of circulation in Ampère-Maxwell’s law.}
\textcolor{black}{
A metacognitive question was given to 65 students from introductory electromagnetism courses at the University of the Basque Country (UPV/EHU) and the University of the Republic (UDELAR), representing all students enrolled in this subject at both institutions. The sample selection took into account that phenomenography aims to include individuals who have experienced the phenomenon in question; therefore, we chose students from the electromagnetism course to whom we had easy access. Additionally, to gain a deeper understanding of students' thought processes, we conducted interviews with 12 students. We used convenience sampling to select students, ensuring data covered the full range of ways in which the phenomenon is experienced. We formed groups of 2 or 3 students, selecting them to represent a wide variety of academic levels and communication skills. This approach aimed to ensure the external validity of the results \cite{MartonPong2005}.}

\textcolor{black}{
All students in the sample had completed the introductory electromagnetism course over 15 weeks, with 4 hours per week dedicated to theory and 2 hours to problem-solving. The content was based on the university physics textbooks by Walker, Resnick and Halliday \cite{walker2014halliday}, Tipler and Mosca \cite{tipler2007physics}, and Young and Freedman \cite{young2019university}, with a week and a half devoted to the study of Ampère-Maxwell’s law. The courses were taught by experienced professors using a lecture-based methodology. In problem-solving sessions, the professor solved problems on the board in front of the group, using exercises from the end-of-chapter sections of the mentioned textbooks.}

\textcolor{black}{Once the question had been prepared, a preliminary test was carried out with students from the Electromagnetism course.} \textcolor{black}{It was confirmed that the students understood the way the proposed solution was written. In addition, the question was validated by consulting six lecturers with extensive experience in teaching electromagnetism. The validation of the interviews followed the same process of initial testing and peer validation.}

The question posed to the students shows a uniform electric field with magnitude increasing with time, confined to a cylindrical region of radius $R$, and states that the magnetic field is to be determined at a point $P$ located at a certain distance from the symmetry axis of the cylindrical region. The step-by-step solution proposed by an \textcolor{black}{imaginary student, Agustina}, to calculate the magnetic field as a function of the parameters of the problem is shown below. \textcolor{black}{This question was selected because it presents a non-standard problem, unlike the examples found at the end of textbook chapters. It prompts students to think critically about how to solve the problem, evaluate the solution strategies provided, and, if necessary, propose alternative approaches.} The proposed solution is correct, but it has the peculiarity that the closed curve chosen by the student exhibits a  semicircle shape. 

\hrulefill

\textbf{Presentation of the metacognitive question}

\begin{figure}[ht!]
\centerline{\includegraphics[width = 0.25\columnwidth]{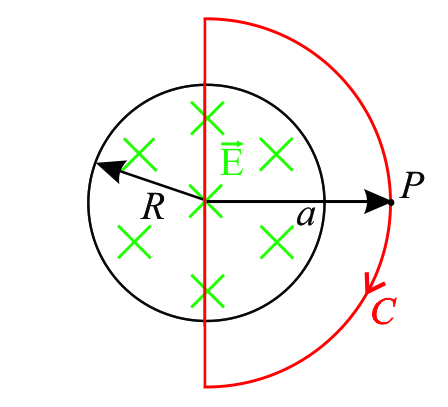}}
\caption{Scheme presented to the students.}
\label{circulación}
\end{figure}

Figure \ref{circulación} shows a uniform electric field which increases with time and is confined to a cylindrical region of radius $R$. Agustina calculates the magnetic field at a point $P$ at a distance \textit{a} from the centre of the cylindrical region.

Analyse the steps that Agustina has taken to solve the problem and comment on each one, explaining why you agree or, if not, what changes you would make.
\vspace{0.2cm}

\textit{Solution:}
To find the magnetic field at $P$, I use a closed curve $C$ in the shape of a semicircle and apply Ampère-Maxwell's law.

\begin{equation}
\oint \vec{B} \cdot \overrightarrow{d l}=\mu_{0} I_{C}+\mu_{0} \varepsilon_{0} \frac{d \Phi_{E}}{d t}
    \end{equation}
Since there are no conduction currents:
\begin{equation}
\oint \vec{B} \cdot \overrightarrow{d l}=\mu_{0} \varepsilon_{0} \frac{d \Phi_{E}}{d t}   
\end{equation}
where $d \Phi_{E} / d t$ is the change in electric field flux in the area enclosed by curve $C$.

Since the magnetic field circulation ( $\oint \vec{B} \cdot \overrightarrow{d l}$ ) along curve $C$ is $B \pi a$.

\begin{equation}
B \pi a=\mu_{0} \varepsilon_{0} \frac{d \Phi_{E}}{d t}
\end{equation}

\begin{equation}
B=\frac{\mu_{0} \varepsilon_{0}}{\pi a} \frac{d \Phi_{E}}{d t}
\end{equation}

\hrulefill
\vspace{0.2cm}

The students were requested to analyse the previous solution and write a comment explaining whether they agree with it or, if not, what changes they would make. The solution proposed to the students is correct and is intended to challenge them in their self-reflection on the validity of the curve for calculating the magnetic field from the magnetic field circulation. Analysing the problem's solution requires students to understand that:
\begin{enumerate}
    \item The Ampère-Maxwell law is valid for any closed curve $C$, regardless of its symmetry.
    \item To calculate the magnetic field circulation, it is necessary to know the shape of the field lines in the context of the defined curve $C$, since its definition is a sum of elements $\vec{B} \cdot \overrightarrow{d l}$.
    \item The Ampère-Maxwell law correlates the magnetic field circulation along the curve $C$ with the net current crossing a surface bounding it, without implying cause-effect relationships \cite{jefimenko2004presenting,Hyodo_2022}.
\end{enumerate}

\textcolor{black}{
The question was posed during the problem-solving class, after the students had already received lessons on the Ampère-Maxwell law. The answers were analysed by two of the researchers using phenomenographic methodology.}
At the beginning of the analysis, each researcher evaluated 20 written tests. A consensus was then reached on the categories of \textcolor{black}{description} that emerged from the students' responses. Each researcher then analysed the remaining tests individually. At the end of this process, the results obtained were compared, and the reliability of the analysis was assessed using Cohen's kappa coefficient, which yielded a value of 0.95, indicating significant agreement \cite{banerjee1999beyond}.

Each interview lasted approximately 20 minutes and in all the cases were conducted by the first author of this paper. The interviewer asked the students to discuss the question aloud and encouraged them to provide detailed explanations of their understanding using non-directive questions and \textcolor{black}{follow-up questions} such as “What do you mean by that?", “Could you explain more?", “Is there anything else you'd like to add about this step?", \textcolor{black}{“What do you think about the election  of the curve $C$ to calculate the magnetic field?" (see \ref{apppendix})}. All interviews were audio-recorded and transcribed. The extracts presented in the results section illustrate the sequence of interpretations and reasoning that characterise the processes of students' development of situated cognition through problem solving. Fictitious names have been used in the presentation of results to ensure anonymity.

\section{\label{RESULTS} Results} 

\textcolor{black}{
According to the phenomenographic analysis, the emerging categories reflect varying degrees of understanding and reasoning of the students when reflecting metacognitively on the resolution of the problem (see Table~\ref{Tabla cuestión}). These categories are qualitatively differentiated according to the level of understanding shown when confronted with the question, standing out for both accuracy in the application of physical concepts and coherence in reasoning, and exhibit a hierarchy of understanding that reflects a progression in the ability to apply concepts coherently. Although the phenomenographic analysis does not focus on individual cases, the examples we provide as empirical justification for the categories necessarily pertain to responses from individual participants.}

\begin{table*}[ht!]
 \caption{\label{Tabla cuestión} \textcolor{black}{Referential and structural aspects and the percentages obtained in each category}.}
\begin{tabular}{c p{6 cm} p{6 cm} c}  
\hline
\textit{Category}
 & \textit{Referential aspects} & \textit{Structural aspects} & \textit{\%}
 \\ \hline
A & \textit{Agreement with the proposed solution.}
Represent the understanding of the magnetic field circulation  and the magnetic field lines. & Divide the curve into segments and analyse the direction of the magnetic field with respect to each segment to determine its contribution to the circulation. & 15.4 \\ 
B & \textit{Disagreement with the calculation of the circulation.}
Refers to a lack of understanding of how certain segments of the curve contribute to the circulation of the field. & They make incorrect assumptions about the contribution of each section of the curve to the flow of the field. & 15.4 \\ 
C & \textit{Disagreement with the choice of the curve}. Express the students' idea that the selected curve should be symmetrical in order to be able to calculate the field. & They use arguments that reflect an incomplete understanding of the role of symmetry in calculating the magnetic field from the AM law. & 20.0 \\
D & \textit{Incoherent answers.}
 & Limited understanding of concepts, without articulation of a coherent strategy. & 15.4 \\ 
E & \textit{No answer}.&Indicates lack of understanding or confidence to address the issue.& 33.8 \\ \hline
\end{tabular}
\end{table*}

\textcolor{black}{
Category A (agreement with analysis), at the top of this hierarchy, groups together answers that correctly interpret the contribution of the magnetic field to the circulation along the selected curve, recognising that the magnetic field in the straight section of the curve does not contribute to the circulation. These students correctly apply the concept of magnetic field circulation, showing an understanding of the shape of magnetic field lines as well as the concept of circulation. The following is an example response in this category (fictitious names are used throughout the text)}.

\begin{quote}
\textit{“Yes, because in $C_{1}$, $\vec{B} \cdot \overrightarrow{d l}=0$ since they are orthogonal throughout $C_{1}$, and in $C_{2}$,  $\vec{B} \cdot$ $\overrightarrow{d l}=B d l$ since they are collinear.”} (\textcolor{black}{Jimena}, see Fig. \ref{Tabla cuestión})
\end{quote}

\begin{figure}[ht!]
\centerline{\includegraphics[width = 0.4\columnwidth]{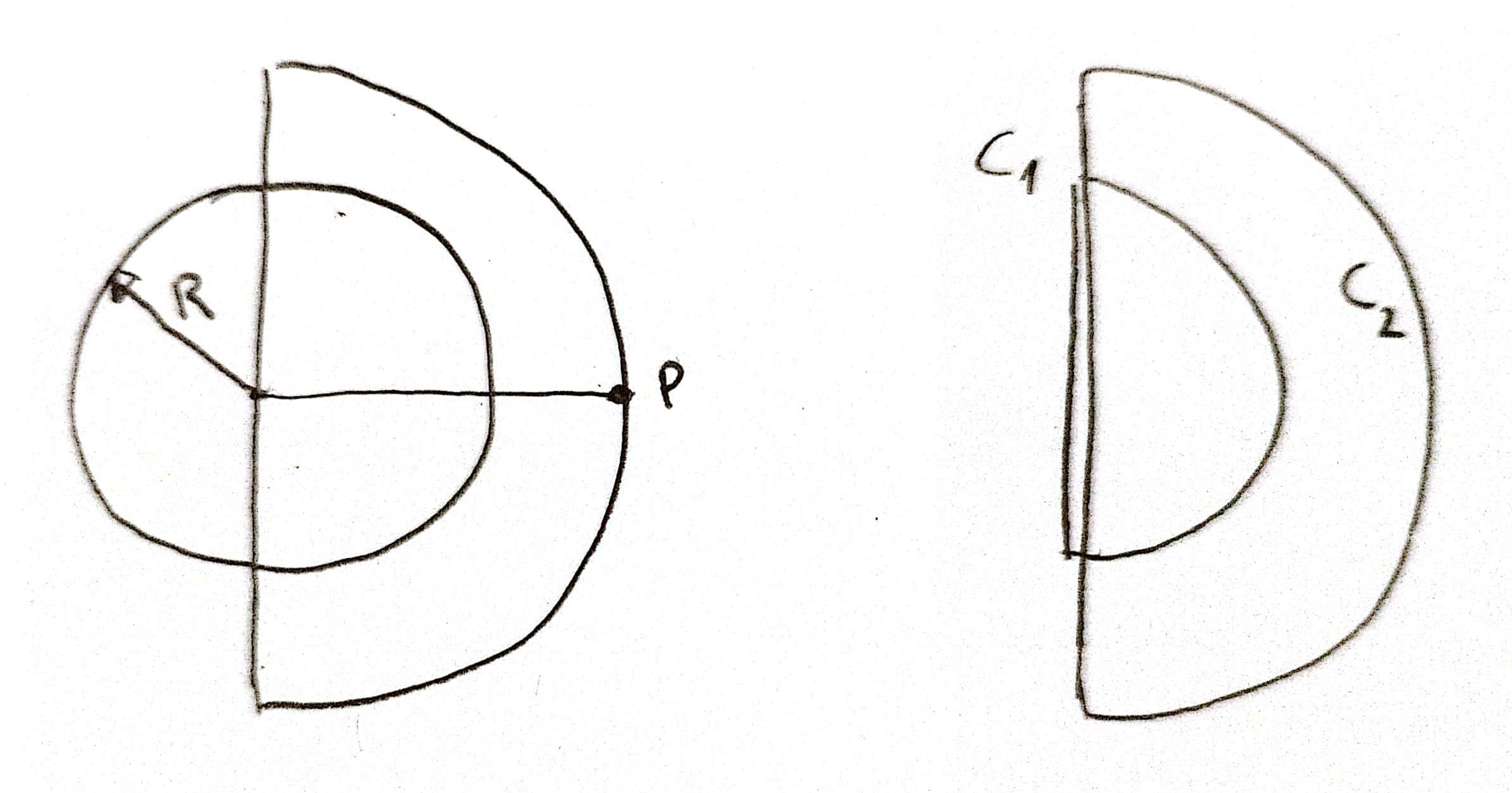}}
\caption{Hand-drawn sketch given by \textcolor{black}{Jimena}.}
\label{cuestion1}
\end{figure}

\textcolor{black}{Jimena} divides the closed curve into two sections, $C_{1}$ and $C_{2}$, and correctly recognises that the contribution to the magnetic field circulation along $C_{1}$ is null. This indicates that, although they do not directly represent the magnetic field along $C_{1}$, they understand that the magnetic field is perpendicular to each element of the curve. On the other hand, the student observes that the magnetic field is collinear with section $C_{2}$. In this way, they demonstrate an adequate understanding of the shape of the magnetic field lines and correctly interpret and apply the concept of magnetic field circulation.

\textcolor{black}{
Category B (disagreement with the calculation of the magnetic field circulation) is at an intermediate level. Students in this category show partial understanding as they consider that the straight part of the curve contributes to the magnetic field circulation. They are not able to see that the magnetic field is perpendicular to the straight line and that its contribution is zero. Here are two examples of answers in this category. }

\begin{quote}
\textit{“dl is $(2+\pi) a$ and the result is $B=\frac{\mu_{0} \varepsilon_{0}}{(2+\pi) a} \frac{d \Phi_{\mathrm{E}}}{d t}$.”}(\textcolor{black}{Marcelo})
\end{quote}

\begin{quote}
\textit{“The induced field B is perpendicular to the field E at every point and its lines are closed. It is not perpendicular to the straight section of curve C, so it should have considered that section in the integral: $B(\pi a+2a)$.”} (\textcolor{black}{Juan})
\end{quote}

\textcolor{black}{Marcelo} believes that the magnetic field circulation should be calculated as the product of the magnetic field and the length of curve $C$. However, they do not recognise that the magnetic field along the straight section of curve $C$ does not contribute to the magnetic field circulation. On the other hand, \textcolor{black}{Juan} considers the direction of the magnetic field, acknowledging its perpendicularity to the electric field, but struggles to use this information effectively to calculate the magnetic field circulation, probably due to an incomplete understanding of circulation.

In the interviews we conducted, we found reasoning that gives us more clues about the different difficulties students exhibit in calculating the magnetic field circulation. In the following dialogue, two students exchange ideas about how to calculate the magnetic field circulation

\begin{figure}[ht!]
\centerline{\includegraphics[width = 0.3\columnwidth]{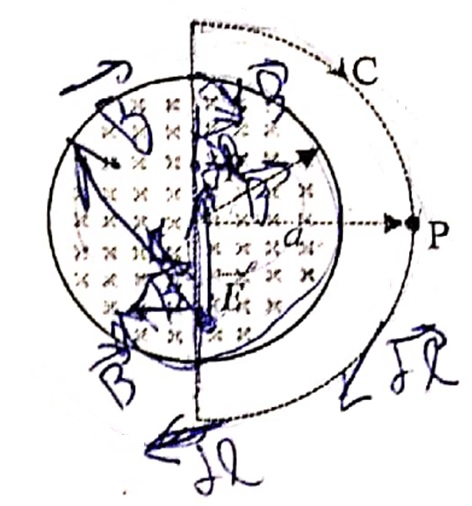}}
\caption{Hand-drawn sketch given by Gustavo during the interview.}
\label{entrevista}
\end{figure}

\begin{quote}
  \textit{Gustavo}: To begin with, it does not take this part into account (in relation to the straight section), because $\pi a$ is only the curved part, not the diagonal.
 \\\textit{Maite}: Sure.\\\textit{Gustavo}: And if you took this diagonal into account (again with respect to the straight section), what would the field look like? Because the field and the \textit{dl} are vectors and the angle has to be taken into account. So, I'm thinking what is this angle on this line, because if it were 90 you can cancel it, I think, if I remember correctly.
I would say that they are circles (referring to magnetic field lines), they cannot be any other shape. It is impossible for it to be any other shape, in here they are circles (referring to the region where the electric field is), yes or yes.
If they are circles, at any point here, you have a field vector like this and a \textit{dl} like this, with an angle less than 90 (draw vectors $\vec{B}$ and $\overrightarrow{d l}$ along the straight section, see Fig. \ref{entrevista}), so there is a magnetic field there and it doesn't take it into account. That's what's wrong. She's not taking that into account.
 \\\textit{Interviewer}: So what do you think is wrong with the solution of the problem?
\\\textit{Gustavo}:  That the integral has a value other than $B\pi a$.
\\\textit{Maite}: Perhaps she is complicating her life by doing this. In other words, maybe if she had chosen the whole circle, it would have been much easier, because she does the circle and that's it and everything is uniform there.
\\\textit{Gustavo}: I think the same.
\end{quote}

Gustavo realises that the magnetic field circulation is a sum of the elementary terms $\vec{B} \cdot \overrightarrow{d l}$ and tries to calculate it by drawing magnetic field lines and $\overrightarrow{d l}$ segments. However, he does not realise that the contribution of the magnetic field to the circulation along the straight section is null because  he has difficulty in correctly representing the magnetic field at different points along this segment.

\textcolor{black}{Category C (disagreement with the choice of curve) reflects a lower level of understanding where students focus on the symmetry of the selected curve. In their answers, they express disagreement about the choice of the curve because, according to them, it should be symmetrical in order to be able to calculate the magnetic field. The following are two examples of responses in this category.}

\begin{quote}
\textit{“No, because in my case the integration curve would be $2 \pi a$ and not $\pi a$.”} (\textcolor{black}{María})
\end{quote}
\begin{quote}
\textit{“I think it is wrong, because the whole surface is not affected by the electric field, you have to consider the whole circle.”} (\textcolor{black}{Ricardo})
\end{quote}

From \textcolor{black}{Maria's answer}, we can deduce that their disagreement with the solution of the question is due to the symmetry of the curve used, considering that a complete circle should be used. On the other hand, \textcolor{black}{Ricardo} argues that a complete circle should be used so that the entire variable electric field crosses the surface bounded by the curve.

The students' answers do not allow us to clearly identify the reasoning underlying their arguments. In the interviews, we found explanations that provide us with additional insights into the types of reasoning employed by students in this category. For example, during the analysis of the questions, one of the interviewees stated:

\begin{quote}
  \textit{Lucía}: The problem is the curve she has chosen. That is, Ampère-Maxwell's law can be used to calculate magnetic fields, but the closed curve must have symmetry. The magnetic field must be the same at all points on the curve, and you also need to know the perimeter.
\end{quote}
As we can see, Lucía clearly explains her reasoning. She states that the calculation of the magnetic field is only possible if the closed curve used is symmetrical. 
\textcolor{black}{In another interview we identified similar reasoning to that of Lucía. In the following dialogue, three students exchange ideas about the choice of the curve C.}

\begin{quote}
\textcolor{black}{
\textit{Florencia}: To find the magnetic field at $P$, take a closed curve $C$ with the shape shown in the figure and apply the Ampère-Maxwell law. The law is correctly written.}

\textcolor{black}{
\textit{Agustín}: Yes. Since there’s no conduction current, ignore the current term.}

\textcolor{black}{
\textit{Emilia}: Now, regarding the choice of the curve, why take that curve? I mean, the worst curve in the world, right? What curve would we choose?}

\textcolor{black}{
\textit{Florencia}: We always go for things that are symmetric. Here, they’re half of symmetric.}

\textcolor{black}{
\textit{Agustín}: Sure, but what happens? You won’t be able to find the magnetic field using the curve you chose, and also because the magnetic field isn’t the same magnitude everywhere. The integral varies, and with that alone, you can’t calculate the field because you can calculate the circulation, but not the field. Do I make sense?}

\textcolor{black}{
\textit{Interviewer}: And you (Emilia and Florencia), what do you think about the choice of curve $C$ for calculating the magnetic field?}

\textcolor{black}{
\textit{Emilia}: It actually works, but you need to know how to do this integral (referring to the line integral of the magnetic field), which will be awful because the magnetic field varies and, besides, the curve isn’t symmetric.}

\textcolor{black}{
\textit{Agustín}: Hold on, what’s the issue? The problem we discussed several times—you’ll be able to calculate the circulation of the field; what you won’t be able to calculate is the field itself.}

\textcolor{black}{
\textit{Emilia}: Yes, but with our mathematical tools. Maybe she knows how to do that.}

\textcolor{black}{
\textit{Agustín}:  What do we always do? What we do here is we say, the modulus of the field is constant here everywhere, so we take it outside and calculate a line integral. But what happens? Here you don't calculate a line integral, you calculate a field integral of I don't know what, I don't know how much, and you calculate the circulation of the field, but you can't...}

\textcolor{black}{
\textit{Florencia}: To solve the equation for $B$.}

\textcolor{black}{
\textit{Agustín}: Sure, you cannot get  $B$.}

\textcolor{black}{
\textit{Emilia}: I agree.}
\end{quote}

\textcolor{black}{
When analysing the curve chosen to determine the magnetic field, Florencia mentions that they always select symmetrical curves, reflecting a strategy learned in class to solve problems of this type. Further on, when discussing the validity of the curve to find the magnetic field, the students argue that it is not possible to calculate it, as due to the shape of the curve, they cannot extract it from the integral. This suggests a limited understanding of the Ampère-Maxwell law, as they fail to interpret the magnetic field circulation as a sum of individual contributions and do not consider the specific shape of the magnetic field.}

Finally, in another interview, we identified an exchange between two students that aligns with what \textcolor{black}{Ricardo} expressed. Below we present an extract from the interview in which the students exchange ideas about the shape of the closed curve.

\begin{quote}
  \textit{Santiago}: Surely (the question) is designed to induce an error, right? But I would consider the other half.
  \\\textit{Interviewer}: What do you mean by the other half? 
  \\\textit{Santiago}: The part of the electric field that we forget.
  \\\textit{Camila}: Of course, you say that this is outside the curve (pointing to the area where there is an electric field that does not cross a surface bounded by $C$)
  \\\textit{Santiago}: Sure, it's only enclosing half of it.
\\\textit{Camila}: It should be a circle. Because it is the shape of the magnetic field generated by this variable electric field.
\\\textit{Santiago}: Right. Because it covers the whole figure that has the electric field (referring to if a complete circle were used as a curve). 
\\\textit{Camila}: Sure, because there is a part of the electric field that you do not take into account, and if you do take it into account, well, we would have to see if it affects it or not.
\\\textit{Santiago}: It affects, it affects, it affects. Maybe the parts that are farther away have less effect (referring to those that are farthest from point $P$), but they still have an effect on the magnetic field.
\end{quote}

Santiago and Camila argue that the magnetic field at point $P$ is due to the entire electric field that crosses the surface bounded by the curve $C$. This idea leads them to conclude that the closed curve $C$ should be a complete circle so that the electric field completely crosses the surface bounded by the curve.

\textcolor{black}{Category D (incoherent) includes answers without a logical structure or with ambiguous reasoning, where students show a limited and confused understanding of magnetic field circulation. At the lowest level of the hierarchy, Category E (no answer) represents no response, suggesting a lack of understanding or confidence in their ability to tackle the problem.}

\section{\label{Discussion}Discussion of results} 

\textcolor{black}{Analysing the descriptive categories in relation to the research question, we find that they reveal different ways of understanding and applying magnetic field circulation, ranging from limited and incomplete interpretations to reflective thinking about the different processes needed to analyse and solve a problem.}

We found that around 15\%  adequately justify why they agree with all the steps in solving the question, showing an understanding of the application of magnetic field circulation in the context of Ampère-Maxwell's law.
We also observed that a similar percentage disagrees with the calculation of the magnetic field circulation, because they do not recognise that the magnetic field does not contribute to the circulation in the straight section. In this sense, we identified reasoning that could be related to the rote application of Ampère-Maxwell's law, where students believe that the magnetic field circulation can always be expressed as the product between the magnetic field and the length of the closed curve. This type of reasoning has been documented in the literature in relation to Ampère's law in different contexts \cite{guisasola2003analisis,guisasola2008gauss,wallace2010}. 

Furthermore, we found that in some cases students do not consider the shape of the magnetic field lines when applying Ampère-Maxwell's law, while in others they do not use the magnetic field information adequately to calculate the magnetic field circulation, nor do they recognise that the magnetic field circulation is a sum (see for example the first interview in section \ref{RESULTS}). 
\textcolor{black}{These findings converge with those already reported} \cite{wallace2010}.

We found that one in five students disagree with the choice of curve for calculating the circulation. Many of them believe that Ampère-Maxwell's law can only be used to calculate magnetic fields if the curve used is symmetrical, which could be due to a functional fixation \cite{viennot1996raisonner} associated with the most used procedure in class, which argues for symmetry to calculate magnetic fields using Ampère-Maxwell's law.

In addition, we found students who disagree with the shape of the curve used in the problem, considering that the variable electric field must cross the entire Amperian curve in order to be able to apply Ampère-Maxwell's law and calculate the magnetic field. \textcolor{black}{Our results suggest that a majority of students have an incomplete understanding of the Ampère-Maxwell law as they do not know that it correlates the magnetic field circulation with the net current that crosses a surface bounding it}. This conceptual difficulty could be due to the belief that Ampère-Maxwell's law establishes cause-effect relationships between the terms involved, in particular with the idea that currents (conduction and displacement) crossing a surface bounded by a curve must be the cause of the magnetic field computed in the line integral. This could lead them to argue that the curve must enclose the entire electric field. Another possible explanation is related to the belief that the magnetic field used in Ampère's law is caused by enclosed currents \cite{guisasola2003analisis,guisasola2008gauss} and its origin could be in the way Maxwell's equations are presented in most introductory physics textbooks, where they are described as cause-effect relationships between their different terms \cite{suarez2023}.

\section{\label{Conclusion}Conclusion and implication for teaching} 

\textcolor{black}{In this article we analysed the different qualitative ways in which students in introductory electromagnetics courses apply the magnetic field circulation in the framework of the Ampère-Maxwell law.}. We designed a metacognitive pencil-and-paper question, presented it in writing to 65 students, and then analysed their answers using phenomenography. We complemented our research by conducting interviews with 12 students.

The circulation of a vector field is an essential part of the language in which Maxwell's laws are expressed. Therefore, understanding and applying it correctly is essential to have a solid learning of the fundamental laws of electromagnetism. Our findings indicate that students still face significant difficulties when applying magnetic field circulation, some of which have been previously reported in relation to Ampère's law. This suggests the complexity inherent in understanding and applying magnetic field circulation among students, regardless of the specific context.

\textcolor{black}{Although the small sample of students limits the generalisability of the study, we believe that the consistency of the results with other studies on Maxwell's laws allows us to draw some implications for the teaching of Ampère-Maxwell's law and magnetic field circulation.  We suggest that in order to promote a deeper understanding of magnetic field circulation in this context and avoid the appearance of reasoning based on functional fixations}, it is crucial to develop teaching-learning sequences that challenge students to calculate the magnetic field from this law, using unconventional curves and solving inverse problems where the magnetic fields cannot be determined from the line integral \cite{Campos2023,Manogue}. This would allow students to reflect on the relationship between the symmetry of the problem, the fields and the circulation. In addition, it is essential to analyse magnetic field lines in different configurations involving conduction and displacement currents, and to apply the Ampère-Maxwell law to different surfaces bounded by the same curve \cite{suarez2022,suarezEJPE}. \textcolor{black}{This would help to recognize that} the Ampère-Maxwell law does not provide information about the sources of the fields, nor does it imply causal relationships, but rather correlations between its different terms \cite{jefimenko2004presenting,Hyodo_2022}.
\textcolor{black}{Finally, non-routinely formulated problems, such as the one presented in this paper, and sense-making activities, such as ‘Ranking Task’ and ‘Conflicting Contentions’, can promote meaningful learning and challenge students' alternative conceptions. \cite{hieggelke2015tipers,Maloney}, can promote scientifically meaningful learning and challenge students' alternative conceptions.}

In our next studies, we will investigate students' conceptual difficulties that could arise from cause-and-effect interpretations of Maxwell's equations, and we will use the results of this work, as well as those of previous research \cite{suarez2023,suarez2024learning}, to design a teaching-learning sequence focused on Ampère-Maxwell's law. This sequence will be implemented in introductory electromagnetism courses and evaluated to determine whether students overcome the conceptual difficulties.

\section*{Acknowledgment}
The research that has been discussed in this publication received funds from the Agencia Nacional de Investigación (ANII), Uruguay, under the code POS\_NAC\_2023\_5\_177677. 
The authors would like to thank PEDECIBA (MEC, UdelaR, Uruguay).  Part of this research was funded by the Spanish government (MINECO\textbackslash FEDER PID2019 -105172RB-I00).

\appendix

\section{\label{apppendix}\textcolor{black}{Protocol
of the interview, design and validation}}

\textcolor{black}{The interview protocol was designed to investigate how students interpret and apply magnetic field circulation in the context of a problem solved by an imaginary student, Agustina. The proposed activity consists of students analysing each step of her solution, in which Agustina chooses a closed semicircular curve to calculate the magnetic field associated with a varying electric field. Working in groups of two or three, students have to verbalise their reasoning aloud during semi-structured interviews and reflect on each step of the resolution. This approach facilitates an exploration of students‘ conceptual difficulties, in line with the phenomenographic approach to capture variations in students’ understanding.}

\textcolor{black}{The protocol is organised in three phases to allow for a continuous and detailed exploration of the students' ideas and reasoning.}

\textcolor{black}{
1. Introduction and initial understanding phase. The aim of this phase is to assess initial understanding of the problem, to identify concepts that students consider relevant and to observe how they perceive the phenomenon collaboratively.
Students are presented with the problem situation: a variable electric field confined to a cylindrical region of radius R and Agustina's intention to determine the magnetic field at a point located at a distance a from the centre of the region.
Opening question: ‘Please have one of you read out the description of the question. Explain in your own words what you think is being asked in the activity’.}

\textcolor{black}{
2. Clarification and reflection phase on the proposed resolution. The aim of this phase is to capture how students experience and understand the problem, identifying conceptual difficulties and ways of reasoning in each step of the resolution. 
Students analyse as a group each step followed by Agustina in her resolution. Reflection is encouraged through non-directive questions and follow-up questions, such as: ‘What do you mean by that?’, ‘Do you agree with this step? Why?’, ‘What would you change in this solution?’, ‘What do you think about the choice of curve C to calculate the magnetic field?’, ‘What do you think about the way the magnetic field circulation is calculated?’. Students discuss their agreements and disagreements and suggest changes if they think any adjustments to the proposed resolution are necessary.}

\textcolor{black}{
3. Justification and conclusions phase. The aim of this phase is to stimulate a comprehensive and shared understanding of the problem, allowing students to express any additional reasoning and reflect collectively on the solution. Students are invited to formulate general conclusions about the steps followed by Agustina, justifying their opinions and consolidating their ideas.
Final question: ‘After analysing the steps followed by Agustina, what do you think as a group about the procedure she used to solve the problem, would you change any aspect or do you think the process correctly reflects what should be done to calculate the magnetic field in this case?}

\textcolor{black}{The design and validation of the protocol was conducted iteratively, while integrating elements of initial validation into design decisions. The initial draft was reviewed by two PER experts to assess the relevance, clarity and ability of the protocol to capture variations in student experiences. This feedback helped to ensure that the questions promoted collaborative analysis without influencing students' reasoning by adjusting the language and clarifying the questions.
Final validation was carried out through a pilot test with two students. The results allowed refining the structure of the protocol, eliminating possible ambiguities and ensuring that both the non-directive and follow-up questions were clear, open-ended and did not interfere with students' collaborative reasoning. This process ensured that the protocol allowed for a wide range of interpretations and reasoning around the application of magnetic field circulation to be observed, aligning with the phenomenographic approach of the study.}

\section*{References} 

\bibliographystyle{iopart-num}

\bibliography{referencias}

\end{document}